\documentclass[amsmath,amssymb,reprint,longbibliography,superscriptaddress]{revtex4-1}
\usepackage{t1enc}
\usepackage[utf8]{inputenc}
\usepackage{amsmath}
\usepackage{graphicx} 
\usepackage{bm}
\usepackage{makeidx}
\usepackage[alsoload=synchem]{siunitx}

\begin{document}
\title{Observation of the $4\pi$-periodic Josephson effect in indium
arsenide nanowires}

\author{Dominique Laroche}
\thanks{These authors contributed equally to this work.}

\author{Dani{\"e}l Bouman}
\thanks{These authors contributed equally to this work.}

\author{David J.~van Woerkom}

\author{Alex Proutski}

\affiliation{QuTech and Kavli Institute of Nanoscience, Delft University of
Technology, 2600 GA Delft, The Netherlands}

\author{Chaitanya Murthy}

\affiliation{Department of Physics, University of California, Santa Barbara, CA
93106, USA}

\author{Dmitry I.~Pikulin}

\affiliation{Station Q, Microsoft Research, Santa Barbara, California
93106-6105, USA}

\author{Chetan Nayak}

\affiliation{Department of Physics, University of California, Santa Barbara, CA
93106, USA}

\affiliation{Station Q, Microsoft Research, Santa Barbara, California
93106-6105, USA}

\author{Ruben J.~J.~van Gulik}

\affiliation{QuTech and Kavli Institute of Nanoscience, Delft University of
Technology, 2600 GA Delft, The Netherlands}

\author{Jesper Nyg{\aa}rd}

\author{Peter Krogstrup}

\affiliation{Center for Quantum Devices and Station Q Copenhagen, Niels Bohr Institute,
University of Copenhagen, Universitetsparken 5, 2100 Copenhagen, Denmark}

\author{Leo P.~Kouwenhoven}

\affiliation{QuTech and Kavli Institute of Nanoscience, Delft University of
Technology, 2600 GA Delft, The Netherlands}

\affiliation{Microsoft Station Q Delft, 2600 GA Delft, The Netherlands}

\author{Attila Geresdi}
\email{To whom correspondence should be addressed;
E-mail:  a.geresdi@tudelft.nl}

\affiliation{QuTech and Kavli Institute of Nanoscience, Delft University of
Technology, 2600 GA Delft, The Netherlands}

\date{\today}

\maketitle

\textbf{Quantum computation by non-Abelian Majorana zero modes (MZMs) offers an
approach to achieve fault tolerance by encoding quantum information in the
non-local charge parity states of semiconductor nanowire networks in the
topological superconductor regime. Thus far, experimental studies of MZMs
chiefly relied on single electron tunneling measurements, which lead to the
decoherence of the quantum information stored in the MZM. As a next step towards
topological quantum computation, charge parity conserving experiments based on
the Josephson effect are required, which can also help exclude suggested
non-topological origins of the zero bias conductance anomaly. Here we report the
direct measurement of the Josephson radiation frequency in indium arsenide
(InAs) nanowires with epitaxial aluminium shells. We observe
the $\mathbf{4\pi}$-periodic Josephson effect above a magnetic field of
$\mathbf{\approx 200\,}$mT, consistent with the estimated and measured 
topological phase transition of similar devices.}

\section*{Introduction}

The universal relation between the frequency $f_\textrm{J}$ of the oscillating
current and an applied DC voltage bias $V$ across a superconducting weak link
\cite{josephson} is determined solely by natural constants:
\begin{equation}
\frac{f_\textrm{J}}{V}=\frac{2e}{h} = \Phi_0^{-1} = 483.6\,\si{\mega
\hertz}\si{\micro\volt}^{-1},
\end{equation}
where $e$ is the single electron charge, $h$ is the Planck constant and $\Phi_0$
is the superconducting flux quantum. This relation, describing the conventional,
$2\pi$-periodic Josephson effect, can be understood as the tunneling of Cooper
pairs with a net charge $e^\star = 2e$ coupled to photons of energy $hf$
\cite{PhysRevLett.106.217005}. This coupling, referred to as the AC Josephson
effect, has first been measured in superconducting tunnel junctions
\cite{Giaever_1965} and has been shown to persist in metallic weak links
\cite{Grimes_1966}, carbon nanotubes \cite{Cleuziou_2007} and semiconductor
channels \cite{Doh_2005, Woerkom_2017-PAT}, as well as in high critical
temperature superconductors \cite{Chen_1987}.

\begin{figure}
\begin{center}
\includegraphics[width=0.45\textwidth]{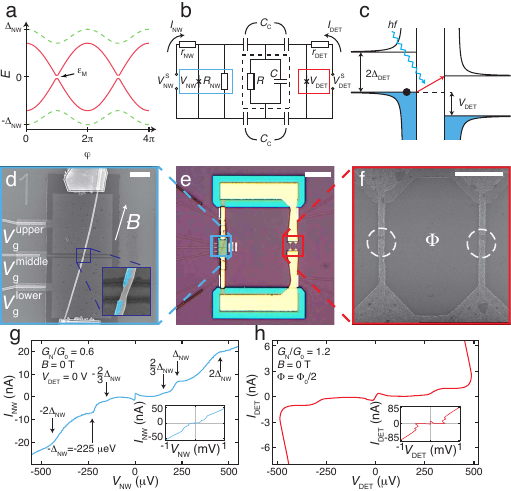}
\caption{\textbf{Principles of the experiment.} \textbf{(a)} 
Energy dispersion of topologically trivial (dashed green line) and nontrivial
(solid red line) Andreev levels inside a NW Josephson junction as a function of
the phase difference across the junction. The gap $\varepsilon_{M}$ arises from
the finite MZMs wavefunction overlap. \textbf{(b)} Equivalent circuit diagram of
the device. The NW junction (in blue box) is capacitively coupled to the
superconducting tunnel junction (red box) via the capacitors $C_{C}$.  The
microwave losses and stray capacitance are modeled by the RC element enclosed by
the dashed black box, see text. The applied DC bias voltages are
$V_{\textrm{NW}}^{\textrm{S}}$ and $V_{\textrm{DET}}^{\textrm{S}}$ with an effective
internal resistance $r_{\textrm{NW}}$ and $r_{\textrm{DET}}$, respectively.
\textbf{(c)} Principle of the frequency sensitive detection based on
photon-assisted tunneling: an absorbed photon with an energy $hf$ gives rise to
quasiparticle current if $hf > 2 \Delta_{\textrm{DET}} - eV_{\textrm{DET}}$.
\textbf{{(d)}} Scanning electron micrograph of the NW junction placed
on three electrostatic gates. A false color micrograph of the junction is shown
in the inset, with the epitaxial Al shell highlighted in cyan. \textbf{(e)}
Bright field optical image of the coupling circuitry between the NW junction
(blue box) and the detector junction (red box). \textbf{(f)} Micrograph of the
split tunnel junction detector. The junctions are encircled. 
\textbf{(g)} Measured $I_{\textrm{NW}}(V_{\textrm{NW}})$ characteristics of the NW
junction at zero in-plane magnetic field exhibiting a supercurrent branch and
multiple Andreev reflections. \textbf{(h)} Measured
$I_{\textrm{DET}}(V_{\textrm{DET}})$ trace of the detector split junction at zero
in-plane magnetic field with a minimized switching current.  The insets in
panels \textbf{(g)} and \textbf{(h)} show the large scale $I(V)$ trace of each
junction. The normal state conductance, $G_\textrm{N}$ is given in the units of
$G_0=2e^2/h$. All images and data were taken on device NW1. The scale bars
denote $1\,\micro$m \textbf{(d)}, $10\,\micro$m \textbf{(e)} and $1\,\micro$m
\textbf{(f)}, respectively.}
\label{Fig1}
\end{center}
\end{figure}

In proximitized semiconductor nanowires, an effective superconducting gap with a
p-wave symmetry arises due to the breaking of the time-reversal symmetry above a
threshold magnetic field \cite{Lutchyn_2010,Oreg_2010, Mourik_2012, Das_2012,
Deng_2012, Churchill_2013, Deng_2016, Zhang_2018}. When a weak link is formed
between two leads, the p-wave component leads to a factor of two increase in the
flux periodicity, giving rise to the so-called $4\pi$-periodic Josephson
effect \cite{doi:10.1063/1.1789931,Fu_2009}. Phenomenologically, this phase
periodicity is equivalent to an effective tunneling charge $e^\star = e$ instead
of $2e$ in Eq.~(1).
Therefore, in this MZM regime, the frequency at a given voltage bias $V$ drops
by a factor of two, $f_\textrm{MZM}(V)=f_\textrm{J}(V)/2$, providing a robust
signature of the topological phase transition in the superconducting leads. In
real devices however, the finite size of the topological regions
\cite{SanJose_2012}, poisoning events \cite{Fu_2009, Lutchyn_2010} and
Landau-Zener tunneling to the quasiparticle continuum \cite{Pikulin_2012-Noise}
can effectively restore the $2\pi$-periodic, trivial state. The latter two
parity-mixing effects cause the system to relax to its ground state, effectively
constraining the system in the lowest topological energy branch (red solid lines
in Fig.~1a). Nevertheless, out-of-equilibrium measurements performed at rates
faster than these equilibration processes can still capture the
$4\pi$-periodic nature of topological junctions \cite{SanJose_2012,
Pikulin_2012-Noise, Houzet_2013}. In contrast, finite-size effects can be
avoided by biasing the junction at voltages large enough to overcome the
Majorana hybridization gap $\varepsilon_\textrm{M}$ \cite{Pikulin_2012-Noise}.

Here, we report the direct observation of a magnetic field-induced halving of
the Josephson radiation frequency \cite{Deblock_2003} in InAs nanowire (NW)
junctions partially covered with an epitaxially grown aluminium shell (Fig.~1d).
In this system, possessing a hard induced superconducting gap
\cite{Krogstrup_2015}, previous direct transport experiments suggest parity
lifetimes above $0.1\,\micro$s \cite{Higginbotham_2015} and hybridization
energies $\varepsilon_\textrm{M} \lesssim 1\,\micro$eV for leads longer than
$1.5\,\micro$m \cite{Albrecht_2016}. Thus, a frequency-sensitive measurement in
the microwave domain is expected to reveal the $4\pi$-periodic Josephson effect
\cite{Rokhinson_2012, Deacon_2017}.

\section*{Results}

\subsection*{Frequency-sensitive detection of the Josephson radiation}
As a frequency-sensitive microwave detector, we utilize a superconducting tunnel
junction with a quasiparticle gap of $\Delta_\textrm{DET}$, wherein the photon-assisted
electron tunneling (PAT) current contributes to the DC current above a voltage
bias threshold $eV_\textrm{DET}>2\Delta_\textrm{DET}-hf$
\cite{Tucker_1985,Deblock_2003} (Fig.~1c).  This \emph{on-chip} detector
\cite{bretheau_nature}, coupled via capacitors $C_\textrm{C}$ to the NW junction
(see Fig.~1b for the schematics and Fig.~1e for an optical image of the device) is
engineered to result in an overdamped microwave environment characterized by a
single $f_\textrm{c}=(2\pi RC)^{-1}\approx 28\,$GHz cutoff frequency with
$R=538\,\si{\ohm}$ and $C=10.4\,$fF, see Supplementary Figure 2. The
resulting broadband coupling to the detector \cite{Woerkom_2017-PAT} inhibits
higher order photon emission, which could mimic the $4\pi$-periodic Josephson
effect \cite{doi:10.1063/1.368113}.

\begin{figure*}
\begin{center}
\includegraphics[width=\textwidth]{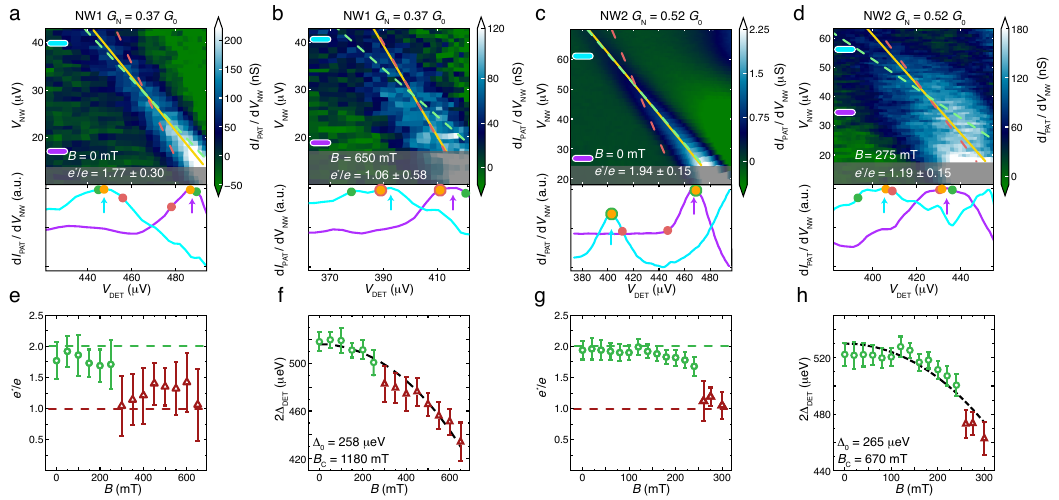}
\caption{\textbf{Magnetic field-induced $\mathbf{4\pi}$-periodic Josephson
radiation.} Differential transconductance
$\textrm{d}{I_\textrm{PAT}}/\textrm{d}{V_\textrm{NW}}$ as a function of $V_\textrm{NW}$ and
$V_\textrm{DET}$ for device NW1 (panels \textbf{(a)} and \textbf{(b)}), NW2
(panels \textbf{(c)} and \textbf{(d)}) at zero and finite magnetic fields,
respectively. The position of the transconductance peak maps the frequency of
the monochromatic Josephson radiation. A linear fit
$e^{\star}V_\textrm{NW}=2\Delta_{\textrm{DET}}-eV_\textrm{DET}$ through these
peaks is shown as an orange line. Dashed green and red lines show linear fits
with a fixed slope corresponding to $e^\star=2e$ and $e^\star=e$, respectively.
The shaded regions show the regimes where the fit of the transconductance peak
is not reliable, see text. Two normalized and smoothed horizontal linecuts are
plotted, where arrows point at the position of the extracted peaks. The orange,
green and red dots denote the position of the best fit, the $e^{\star} = 2e$ fit
and the $e^{\star} = e$ fit, respectively.  The evolution of $e^\star(B)$ and
$2\Delta_\textrm{DET}(B)$ are presented in panels \textbf{(e)} and \textbf{(f)}
for NW1 and in panels \textbf{(g)} and \textbf{(h)} for NW2. For the calculation
of the error bars, see text and Supplementary Note 2. The transition from the
$2\pi$- to $4\pi$-periodic Josephson radiation is observed between $175$ and
$300\,\si{\milli \tesla}$ as $e^\star$ evolves from values near $2e$ (green
circles) to values close to $1e$ (red triangles). For all devices,
$2\Delta_\textrm{DET}(B)$ drops monotonically (black dashed line, see text),
independently of the change in $e^\star$.}
\label{Fig2}
\end{center}
\end{figure*}

The nanowire is deterministically deposited on a set of three gates covered by
$30\,$nm thick SiN$_x$ dielectric as shown in Fig.~1d. The Josephson weak link,
where the Al shell is removed by wet chemical etching, is located above the
central gate (see inset of Fig.~1d). We investigated devices with junction
lengths ranging from $86\,$nm to $271\,$nm. The high quality of the nanowire
junction is apparent from the presence of distinct multiple Andreev reflection
steps in its $I_\textrm{NW}(V_\textrm{NW})$ characteristics \cite{Scheer_1997}
(Fig.~1g), which is a signature of the hard superconducting gap in the nanowire
\cite{Krogstrup_2015}. The observed curves and linear conductance also establish
that no conductive mode with a transmission close to unity exists in the
channel, which could contribute to the $4\pi$-periodic signal even in the
absence of topological ordering \cite{Pikulin_2012-Noise}.

The microwave detector, presented in Fig.~1f, is fabricated using two
angle-evaporated \cite{Dolan_1977} Al/AlO$_x$/Al tunnel junctions, forming a
superconducting quantum interference device (SQUID). This geometry allows us to
minimize the Josephson energy of the detector by applying $\Phi=\Phi_0/2$ flux
through the loop (see Fig.~1h) and thus to limit its backaction to the nanowire.
The respectively $8$ and $11\,$nm thick Al layers set an \emph{in-plane}
critical magnetic field of the detector in excess of $1\,$T, well above the
measured topological transition in similar devices \cite{Albrecht_2016}.
Nevertheless, increasing subgap currents limited our experimental field range
to $325-650\,$mT for different devices. The circuit parameters and
fabrication details are given in the Supplementary Tables and in the Methods,
respectively.
         
In the presence of a voltage spectral density $S_{V}(f)$, the DC
current contribution of the PAT process is as follows \cite{Tucker_1985,
Deblock_2003} in the subgap regime, where $eV_\textrm{DET} <
2\Delta_\textrm{DET}$:
\begin{equation}
\label{eq2}
I_\textrm{PAT}(V_\textrm{DET}) = \int_{0}^{\infty} \textrm{d} f \left(
\frac{e}{hf}\right)^{2} S_{V}(f) I_{\textrm{QP},0}\left( V_\textrm{DET} +
\frac{hf}{e}\right).
\end{equation}
Here, $I_{QP,0}(V_\textrm{DET})$ is the tunnel junction current in the absence
of absorbed radiation, $S_V (f)=0$ (see Fig.~1h). Note that the quasiparticle
gap edge at $eV_\textrm{DET} = 2\Delta_\textrm{DET}$ results in a sharp
increase of $I_{QP,0}(V_{\textrm{DET}})$. In the presence of monochromatic
radiation with a frequency $f_0$, $S_V(f)\sim \delta(f-f_0)$,
$I_\textrm{PAT}(V_\textrm{DET})$ thus develops a step-like feature at $hf_0 =
2\Delta_\textrm{DET} - eV_\textrm{DET}$. With a phenomenological effective
charge $e^{\star}$ of the AC Josephson effect, we write this condition in terms
of the voltage drop on the nanowire, $V_\textrm{NW}$:
\begin{equation}
\label{eq3}
e^{\star}V_\textrm{NW}= h f_0 = 2\Delta_\textrm{DET}-eV_\textrm{DET},
\end{equation}
where $e^{\star} = 2e$ for conventional junctions (see Eq.~(1)) and $e^{\star} =
e$ in the $4\pi$-periodic regime. To extract $e^{\star}$ and thus determine the
periodicity of the Josephson radiation, we track the transconductance peak
$\textrm{d} I_\textrm{PAT}/\textrm{d} V_\textrm{NW}(V_\textrm{NW},
V_\textrm{DET})$ measured by standard lock-in techniques at a frequency of
$17.7\,$Hz (see Supplementary Figure 1). The experiments were performed at the
base temperature of a dilution refrigerator ($\sim 20\,$mK).

\subsection*{The Josephson radiation as a function of the magnetic field}

\begin{figure*}
\begin{center}
\includegraphics[width=\textwidth]{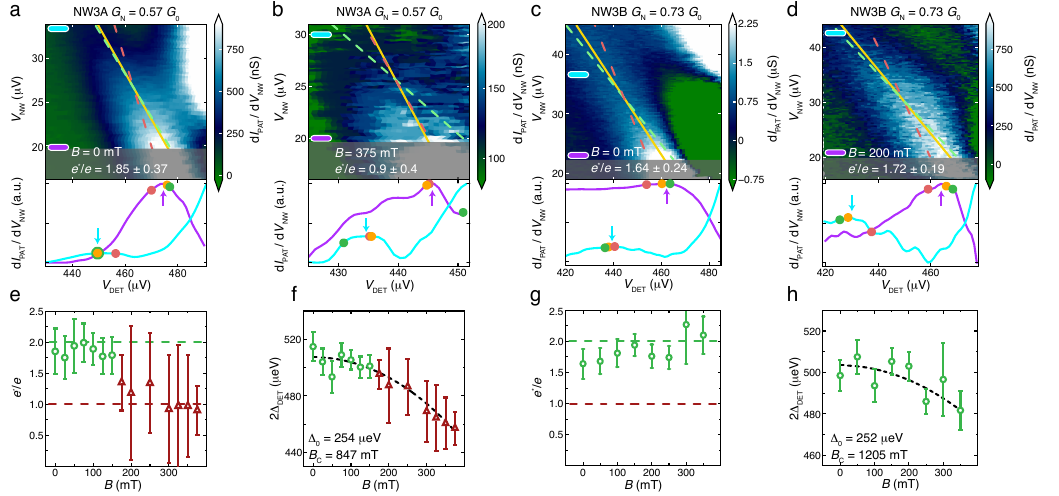}
\caption{\textbf{Gate tuning of the $\mathbf{4\pi}$-periodic radiation regime.}
Differential transconductance $\textrm{d}{I_\textrm{PAT}}/\textrm{d}{V_\textrm{NW}}$ as a
function of $V_\textrm{NW}$ and $V_\textrm{DET}$ for device NW3 at gate setting
A (panels \textbf{(a)} and \textbf{(b)}) and setting B (panels \textbf{(c)} and
\textbf{(d)}) at zero and finite magnetic field, respectively. A linear fit and
fits with a fixed slope $e^\star=2e$ and $e^\star=e$ are shown as an orange
line, a dashed green line and a dashed red line, respectively. Two normalized
and smoothed horizontal linecuts are also presented, where arrows point at the
position of the extracted peaks. The evolution of $e^\star(B)$ and
$\Delta_\textrm{DET}(B)$ are shown in panels \textbf{(e)} and \textbf{(f)}
for setting A and in panels \textbf{(g)} and \textbf{(h)} for setting B. A
transition from to $2\pi$- to $4\pi$-periodic Josephson radiation is observed
for gate setting A, but the radiation remains $2\pi$-periodic for setting B. The
gate voltage values are shown in Supplementary Table 2. For the calculation
of the error bars, see text and Supplementary Note 2.}
\label{Fig3}
\end{center}
\end{figure*}

Typical experimental datasets are shown in Fig.~2 for two nanowire junctions
(NW1 and NW2, respectively) as the source of Josephson radiation. We limit the
detector voltage range by the condition $\textrm{d} I_\textrm{DET}/\textrm{d} V_\textrm{DET} <
10\,\micro$S where the subgap quasiparticle current is still negligible,
typically $I_\textrm{DET} \lesssim 1\,$nA. A lower limit of the emitter junction
voltage is defined by the phase diffusion regime \cite{Ivanchenko_1969},
characterized by periodic switching and retrapping events, which breaks the
validity of Eq.~(1) (see Supplementary Note 3). We therefore do not consider the
low $V_\textrm{NW}$ regime, within the supercurrent peak. We show this range,
excluded from the linear fits, shaded in grey in Fig.~2 and Fig.~3 (see
Supplementary Note 2 on the characterization of these limits). We fit the peak
positions using Eq.~(3) in order to extract $e^\star$ and $\Delta_\textrm{DET}$
as a function of the applied in-plane magnetic field. The typical standard
deviation is $3.6\,$GHz for each frequency datapoint (see Supplementary Note 2).
The error bars of the fitted parameters are determined using the bootstrapping
method \cite{10.2307/2245500} (see Supplementary Note 2) and show the full width at
half maximum yielding a confidence level of $75\%$ for a Gaussian lineshape.

At zero magnetic field (Fig.~2a and c), the
emitted Josephson radiation is always $2\pi$-periodic with an extracted effective charge
close to $e^{\star} = 2e$, as shown by the good agreement between the orange
line and the dashed green line (best fit with fixed $e^{\star} = 2e$). In
contrast, NW1 and NW2 exhibit the $4\pi$-periodic Josephson effect above a
threshold magnetic field (Fig.~2b and d), where $e^{\star} \approx e$.
The full evolution is shown in Fig.~2e and 2g, respectively, where a sharp
transition is visible from $e^{\star} \approx 2e$ (green circles) to $e^{\star}
\approx e$ (red triangles). Finally, the fitted $\Delta_\textrm{DET}$
(Fig.~2c and f) shows a monotonic decrease described by
$2\Delta_\textrm{DET}(B)=2\Delta_{0}\sqrt{1-B^2/B_\textrm{c}^2}$ for all devices
(dashed lines), with no additional feature at the transition field. In contrast with the
nanowire junctions, our control device, an Al/AlO$_x$/Al tunnel junction,
exhibits no transition in $e^{\star}$ over the entire magnetic
field range (see Supplementary Figure 5).

\subsection*{The Josephson radiation at different gate voltages}

\begin{figure*}
\begin{center}
\includegraphics[width=\textwidth]{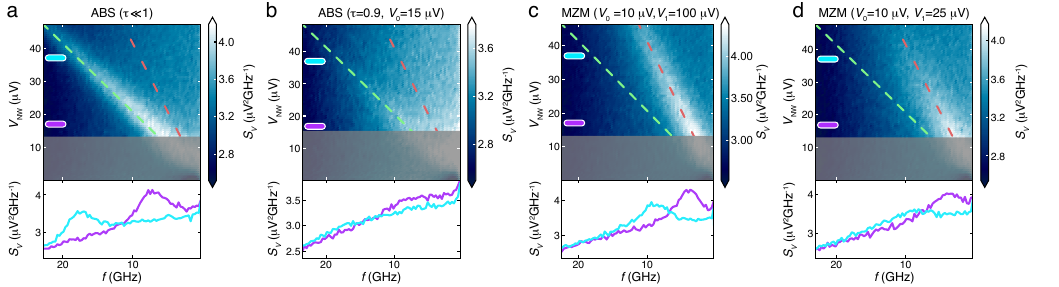}
\caption{\textbf{The calculated radiation spectrum.}
The voltage spectral density $S_{V}(f)$ incident on the detector junction,
computed by numerically solving the system of stochastic differential equations
shown in Supplementary Note 3.
Panels \textbf{(a)} and \textbf{(b)} show results for a junction in the trivial
regime (small transmission and large transmission, respectively), while panels
\textbf{(c)} and \textbf{(d)} show the emission spectrum in the topological
regime. $V_{0}$ and $V_{1}$ are voltage scales for Landau-Zener tunneling
between branches of the junction bound state and for tunneling to the quasiparticle
continuum, respectively, see text. Circuit parameters are set as
$r_{\textrm{NW}} = 2.4 \, \si{\kilo\ohm}$, $R_{\textrm{NW}} = 50\,\si{\kilo\ohm}$, $R = 0.5
\,\si{\kilo\ohm}$, $C = 10 \,\si{\femto\farad}$, $C_{C} = 400 \,\si{\femto
\farad}$, and $I^{0}_{\textrm{NW}} = 8 \,\si{\nano\ampere}$. The noise temperature
is $T = 150\,$mK and the quasiparticle poisoning rate is $\Gamma_{q} = 100
\,\si{\mega\hertz}$. As in Fig.~2, the dashed green and red lines show the
frequency of the Josephson radiation corresponding to $e^{\star} = 2e$ and
$e^{\star} = e$, respectively. The estimated phase diffusion region is shaded in
gray.}
\label{Fig4}
\end{center}
\end{figure*}

Fig.~3 shows the magnetic field evolution of device NW3 at two distinct gate
settings with similar $G_\textrm{N}$ and $\textrm{d} I_\textrm{PAT}/\textrm{d} V_\textrm{NW}$
corresponding to similar Josephson coupling. By tuning the chemical potential in
the nanowire via changing the gate voltages, it is possible to displace the position of the onset of the
$4\pi$-periodic Josephson radiation from $\approx 175 \, \si{\milli \tesla}$
(Fig.~3b) to values larger than $375\, \si{\milli \tesla}$ (Fig.~3d). Note that
the additional local maximum at high $V_\textrm{NW}$ values, also observed in
earlier experiments \cite{Woerkom_2017-PAT}, is attributed to the shot noise of
the nanowire junction.

The possibility to tune the nanowire devices into the $4\pi$-periodic
Josephson radiation regime with both magnetic field and chemical potential is
consistent with the predicted phase diagram of this system
\cite{Fu_2009,Lutchyn_2010,Oreg_2010}. We observe the same behaviour in four
distinct nanowire devices (see Supplementary Figure 4 for device NW4), which we
can interpret within the single subband model of the topological phase transition that takes
place at a magnetic field $B^\star$, where $E_\textrm{z} = g \mu_\textrm{B}
B^\star / 2 = \sqrt{\Delta_\textrm{NW}^{2} + \mu_\textrm{NW}^{2}}$.
Here $g$ and $\mu_\textrm{B}$ are the Land\'e g-factor and the Bohr magnetron,
respectively. From our device parameters (see Supplementary Table 2), lower
bounds on the g-factors ranging from $g \approx 11$ ($B^{\star}= 175\,
\si{\milli \tesla}$) in device NW3 to $g \approx 35$ ($B^{\star}= 190\,
\si{\milli \tesla}$) in device NW4 are obtained, in agreement with values
reported in similar devices \cite{Deng_2016, Albrecht_2016, Woerkom_2017-ABS}.
In contrast, an accidental crossing of a trivial Andreev bound state would be
inconsistent with the observed field range of $\Delta B \sim 0.3\,$T of the
$4\pi$-periodic radiation, since within this range, a spinful Andreev level
\cite{Woerkom_2017-ABS} would evolve over the scale of the superconducting gap,
$\Delta_\textrm{NW} \sim g \mu_\textrm{B} \Delta B$, suppressing the $4\pi$
periodicity. We however did not observe a continuous variation of
the onset magnetic field $B^\star$ as a function of the applied gate voltages.
This behaviour is consistent with calculations of the topological
phase diagram based on realistic device simulations including orbital effects of
the magnetic field \cite{PhysRevB.93.235434} and multiple spatial dimensions
\cite{PhysRevLett.105.227003, PhysRevB.95.115421} of the device.

We observe a single Josephson radiation frequency  in the
$4\pi$-periodic regime, which is consistent with the supercurrent being
predominantly carried by a single transmitting mode. While we were not able to
reliably extract the transparency and the number of modes in our devices, the
single mode regime was observed earlier in similar InAs nanowires
\cite{Woerkom_2017-ABS, Spanton_2017, Goffman_2017}. We also note that an
upper bound on the channel transmission of $\tau = G_\textrm{N} / G_0$ can be
determined from the normal state conductance $G_\textrm{N} < G_0$, which is
measured in the linear regime well above the superconducting gap. This value is shown in
Fig.~2 and Fig.~3 for each device.

\subsection*{Numerical simulations of the Josephson radiation frequency}

Next, we numerically evaluate the expected voltage spectral density seen by the
detector junction in various regimes. We use the quasiclassical resistively and
capacitively shunted junction (RCSJ) model coupled to a stochastic differential
equation describing the occupation of the single pair of Andreev levels in the
NW junction. The equivalent circuit of the device in the microwave domain is
shown in Fig.~1b, where each element is experimentally characterized
\cite{Woerkom_2017-PAT} (see Supplementary Figure 2 and Supplementary Tables).
Note that we neglect the load of the detector on the circuit, which is justified by its negligible subgap
conductance compared to that of all other elements in the circuit.

Our model of the nanowire junction considers Landau-Zener (LZ) tunneling between
branches of the energy-phase dispersion shown in Fig.~1a, as well as tunneling
to the continuum, and stochastic quasiparticle poisoning events
\cite{Pikulin_2012-Noise}. The probability of LZ tunneling is determined by the voltage drop
$V_\textrm{NW}$ according to
$P_\textrm{LZ}=\exp(-V_0/V_\textrm{NW})$, where $eV_0 = 4\pi
\varepsilon_\textrm{M}^2/(\Delta_\textrm{NW} \sqrt{\tau})$ is the characteristic voltage
above which $P_\textrm{LZ}\sim 1$. In this limit,
$4\pi$-periodicity is observed despite
the gap $\varepsilon_\textrm{M}$ caused by finite-size effects \cite{Albrecht_2016}.
Similarly, LZ tunneling to the continuum close to $\varphi=2\pi$ defines a
voltage scale $eV_1 = 2\pi \Delta_\textrm{NW} (1-\sqrt{\tau})^2/\sqrt{\tau}$,
above which $2\pi$-periodicity is restored \cite{Pikulin_2012-Noise}. We
note that a trivial Andreev bound state in the short junction limit
can be modeled similarly with $e V_0 = \pi \Delta_\textrm{NW} (1-\tau)$ and
$eV_1 = 0$.

Fig.~4 shows representative plots obtained by numerically evaluating $S_V(f,
V_\textrm{NW})$ (see Supplementary Note 3), which determines the photon-assisted
tunneling current by Eq.~(2). We observe that the numerical results agree well
with the characteristic features of the experimental data. We find that the
circuit equations allow for a phase diffusion regime at low $V_\textrm{NW}$
values \cite{Ivanchenko_1969}, where $e^\star V_\textrm{NW} < hf$, because the
junction spends part of the time in the steady supercurrent state where the
voltage drop is zero. The calculations also reproduce the absence of higher
harmonics in the radiation spectrum, attributed to the low transmission of the
junction and the overdamped nature of the microwave environment
\cite{doi:10.1063/1.368113}. This confirms our expectation of the suppression of
multiphoton processes due to a low quality factor, justifying the usage of the
semiclassical junction model.

A key result of these simulations in a wide range of junction parameters is
that, with the circuit elements taking values representative of those in the
experiment, the radiation frequency always reflects the internal dynamics of the
nanowire Josephson junction both in the $2\pi$-periodic (Fig.~4a and b) and in
the $4\pi$-periodic emission regime (Fig.~4c and d). Finally, we note that our
results are consistent with $V_0 \lesssim 15\,\micro$eV translating to an
avoided crossing $\varepsilon_\textrm{M} \lesssim 10 \,\micro$eV. Using the
exponential cutoff in Ref.~\cite{Albrecht_2016}, this suggests that our devices have a
continuous topological region of several hundreds of nanometers on each side of
the nanowire junction, which is consistent with the scanning electron microscopy
images of the devices.

\section*{Discussion}

In conclusion, we observed the $4\pi$-periodic Josephson effect in multiple InAs
nanowires above a threshold magnetic field in a range of $175-300\,$mT. This
effect, which can be suppressed by tuning the gate voltages, is consistent with
the expected signatures of a topological phase transition. By observing the
periodicity of the Josephson effect using an \emph{on-chip} microwave detector,
we investigated this system whilst preserving its charge parity, in line with
the requirements for prospective topological quantum computers. This
experimental technique may also prove instrumental in identifying more exotic
non-Abelian anyon states \cite{PhysRevX.2.041002, Clarke_2013}, due to its
proven sensitivity to the periodicity of the Josephson effect, directly
measuring the charge fractionalization of the anyon state
\cite{PhysRevLett.113.036401, Laflamme_2016}.

\section*{Methods}

\subsection*{Device Fabrication}

The devices are fabricated on commercially available undoped Si substrates with
a $285\,$nm thick insulating SiO$_x$ layer in a similar fashion to Refs.
\cite{Woerkom_2017-ABS} and \cite{Woerkom_2017-PAT}.  All etching and metal
deposition steps are realized using standard positive tone electron-beam
lithography techniques.  First, three Ti/Au ($5\,$nm/$15\,$nm) electrostatic
gates and the coupling capacitor bottom plates are deposited (see Fig.~1 for design
details).  These are subsequently covered by a $\sim 30\,$nm thick SiN$_x$
dielectric layer deposited by sputtering.  Eleven $100\,$nm wide Cr/Pt
($5\,$nm/$25\,$nm) tracks are then defined.  These $\sim 100\, \si{\ohm \micro
\meter}^{-1}$ resistive lines connect the gates, the (yet to be
defined) Al/AlO$_x$/Al detector and the nanowire to the instrumentation setup.
Next, the Al/AlO$_x$/Al Josephson junctions are fabricated by evaporating $8$
and $11\,$nm thick Al layers with an intermediate \emph{in-situ} oxidation step
at $0.5\,$mbar for $4\,$minutes using the Dolan bridge technique
\cite{Dolan_1977}.
The nanowires are then deterministically deposited onto the electrostatic gates with
a micro-manipulator setup equipped with an optical microscope.  A gap in the
nanowire Al shell is then created by wet etching using Transene D at a
temperature of $48.2^{\circ}$C  for $12\,$seconds.
Next, both the nanowire and the detector junctions are connected to the
resistive lines with an $80\,$nm thick sputtered NbTiN film after an
\emph{in-situ} Ar plasma milling step.  Finally, a Ti/Au ($15/100\,$nm) layer is
evaporated to define quasiparticles traps, the upper capacitor plates and the contact pads.  We note
that no NbTiN film was used in device NW3.  Instead, a Ti/Au ($15/100\,$nm)
layer was used to contact the nanowire and the detector. The dimensions and properties
of each device are presented in Supplementary Table~1, and
the experimental setup is described in Supplementary Figure~1.  We
note that the detector is made of narrow and thin aluminum sections (see Fig.~1f) to limit the presence of
vortices near the Al/AlO$_x$/Al junctions, and thus to decrease the subgap current
in finite magnetic field.
   
The InAs nanowires used in this work are grown via a two-step process by
molecular beam epitaxy. The InAs nanowires are grown at $420^{\circ}$C using
the vapour-liquid-solid method with Au droplets as a catalyst. After cooling the
system to $-30^{\circ}$C, Al is epitaxially grown on two of the six nanowires
facets \cite{Krogstrup_2015}.

\subsection*{The microwave environment of the InAs Josephson junction}

We model the effective microwave environment of the nanowire Josephson junction
with a parallel lumped resistor (R) and capacitor (C) element, which accounts for the low-pass nature of the
coupling circuit (see inset of Supplementary Figure 2a). We determine the
effective RC values by measuring a sample wherein the nanowire junction is
replaced by an Al/AlO$_x$/Al tunnel junction. The supercurrent peak is fitted
against the Ivanchenko-Zil'berman model to find the RC values and the noise
temperature of the circuit \cite{Ivanchenko_1969} at zero magnetic field (see
Supplementary Figure 2a). The critical current as a function of the magnetic
field is then found using the same model, with the R,C and the noise temperature
fixed at their zero field value (Supplementary Figure 2b). We note that the same
coupling circuit was used in \cite{Woerkom_2017-PAT}, leading to RC and noise temperature values in good
agreement with the current ones.  Thus, we conclude that the reproducibility is
good for all samples featured in the current study. These parameters are used to
theoretically study the dynamics of the Josephson radiation.

\subsection*{Reproducibility of the transition for nanowire devices}
Supplementary Figure 3 shows every differential transconductance
color plots from which the effective charge $e^{\star}$ has been extracted in Fig.~2e. 
The color plots nominally follow the same trend as the ones presented in Fig.~2.
 Supplementary Figure 4 shows the magnetic field evolution of $e^{\star}$ in
 device NW4. Device NW4 also exhibits a transition from to $2\pi$- to
$4\pi$-periodic Josephson radiation at $B \sim 175 \,$mT.  As such, the
observation of a magnetic field induced transition in the periodicity of the
Josephson radiation has been observed in four distinct devices, showcasing the
reproducibility of the observation.

\subsection*{Josephson radiation of an Al/AlO$_x$/Al tunnel junction}
Supplementary Figure 5 shows our measured data with a conventional
Al/AlO$_x$/Al superconducting tunnel junction as the source of Josephson
radiation. Evaluating $e^\star$ as a function of magnetic field in the same
range as for Fig.~2 and Supplementary Figure 4, we observe no transition in the
periodicity of the Josephson effect, confirming that the $4\pi$-periodic
Josephson radiation only occurs in nanowire junctions. We note that, in
order to keep the circuit behavior similar, the normal state conductance of the
tunnel junction was set to $G_\textrm{N,T}=0.26 G_0$.

\section*{Data availability}

The datasets analysed during this study are available at the
4TU.ResearchData repository, DOI:
10.4121/uuid:1f936840-5bc2-40ca-8c32-1797c12cacb1 (Ref.~\cite{rawdata}).

\section*{Acknowledgements}

The authors acknowledge O.~Benningshof, J.~Mensingh, M.~Quintero-P\`{e}rez and
R.~Schouten for technical assistance. This work has been supported by the Dutch
Organization for Fundamental Research on Matter (FOM), Microsoft Corporation
Station Q, the Danish National Research Foundation and a Synergy Grant of the
European Research Council.
A.~G.~acknowledges the support of the Netherlands Organization for Scientific
Research (NWO) by a Veni grant.

\section*{Author Contributions}

D.~L., D.~B., D.~J.~v.~W. and A.~P.~fabricated the samples and performed the
experiments. P.~K.~and J.~N.~contributed to the nanowire growth. L.~P.~K.~and
A.~G.~designed and supervised the experiments. C.~M., D.~P.~and C.~N.~developed
the theoretical model of the devices. D.~L., D.~B., D.~J.~v.~W.,
R.~J.~J.~v.~G., L.~P.~K.~and A.~G.~analyzed the data. The manuscript has been
prepared with contributions from all the authors.

\section*{Competing Interests}

The authors declare no competing interests.

\bibliographystyle{naturemag}
\bibliography{bibliography4pi}

\end{document}